\documentclass[a4paper,10pt]{article}
\usepackage{graphics}
\begin{document}
\begin{center}
 {\bfseries Separation of Flip and Non-Flip parts\\
 of $np\to pn$ Charge Exchange at energies T$_n = 0.5-2.0$\,GeV}
 \vskip 5mm
 \underline{R.A.~Shindin}$^*$, E.V.~Chernykh, D.K.~Guriev, A.A.~Morozov,\\
 A.A.~Nomofilov, V.Yu.~Prytkov, V.I.~Sharov, L.N.~Strunov
 \vskip 5mm {\small
 {\it JINR, Veksler and Baldin Laboratory of High Energies, 141980 Dubna, Russia}\\
 $*\quad$ {\it E-mail: shindin@sunhe.jinr.ru}}
\end{center}

\vskip 5mm

\begin{abstract}
 The new Delta-Sigma experimental data on the ratio $R_{dp}$
 allowed separating the Flip and
 Non-Flip parts of the differential cross section
 of $np\to pn$ charge exchange process
 at the zero angle by the Dean formula.
 The PSA solutions for the $np\to np$
 elastic scattering are transformed to
 the $np\to pn$ charge exchange representation
 using unitary transition,
 and good agreement is obtain.
\end{abstract}

 \section{Introduction}
 The Delta-Sigma experiment research program \cite{bib1} intends
 to obtain a complete $np$ data set at the zero angle: 
 the measurements of total cross section differences
 $\Delta\sigma_{\rm L}(np)$ and $\Delta\sigma_{\rm T}(np)$
 for the longitudinal ($\rm L$) or transverse ($\rm T$) beam and target
 polarizations and spin-correlation parameters
 $A_{00kk}(np)$ and $A_{00nn}(np)$ \cite{bib2}
 as well as unpolarized measurements of values $\sigma_{0\,tot}(np)$,
 $d\sigma/dt(np\to pn)$ and $R_{dp}$.
 The main task of these studies is to determine the {\it Re} and {\it Im}
 parts of $np$ amplitudes over the energy region 1.2--3.7\,GeV.
 The energy dependence of $\Delta\sigma_{\rm L}(np)$ \cite{bib2} shows an anomalous
 decrease to zero above 1.2\,GeV and the structure
 in $-\Delta\sigma_{\rm L}(\rm{I=0})$ around 1.8\,GeV \cite{Strun}
 predicted in \cite{Bag-Model 1,Bag-Model 2}.
 For the exhaustive analysis of this structure it is necessary to build the Argand diagrams
 for the {\it Re} and {\it Im} parts of each of the three $NN$ forward scattering amplitudes.
 To reduce the sign ambiguities the
 Delta-Sigma collaboration measured the
 ratio $R_{dp}=d\sigma/dt(nd)\bigm/d\sigma/dt(np)$ for
 the charge exchange quasi-elastic and elastic processes at $0\,^\circ$
 using the D$_2$ and H$_2$ targets.
 The knowledge of $R_{dp}$ could provide
 additional constraint and will allow one of some
 sign uncertainties to be eliminated for the direct reconstruction
 of the {\it Re} parts of the scattering amplitudes.

 The $R_{dp}$ value at zero angle 
 defines the ratio $r^{\rm nfl/fl}$ of the
 Non-Flip to Flip contributions in the $np\to pn$
 charge exchange process.
 This possibility is based on the deuteron properties
 that the deuteron is the
 amplitude filter at small momentum transfer
 in the $nd\to p(nn)$ reaction, and the Non-Flip part
 vanishes due to the Pauli principle
 for two slow neutrons.
 Therefore, the quasi-elastic $nd$ differential cross section
 is the Flip yield of the $np\to pn$ charge exchange process.
 It is expressed by the Dean formula
 \cite{Dean-1,Dean-2,Luboshitz to Dean}.

 The $np$ elastic reaction can be represented
 by two approaches: either as
 the charge exchange $np\to pn$ reaction
 to the $\theta$ angle ($\theta=\theta_{\rm CM}$) or
 as the neutron elastic scattering $np\to np$
 in the inverted direction to the $(\pi-\theta)$ angle.
 Though both representations
 have equivalent differential
 cross sections, their Flip or Non-Flip
 parts are absolutely different \cite{LL 1,LL 2}.
 The main cause for this distinction will be
 shown in section 3.
 To compare the energy dependencies of experimental $R_{dp}$ or
 estimated $r^{\rm nfl/fl}$ with the PSA solutions
 of $np$ elastic scattering,
 we should use the true charge exchange amplitudes,
 which requires the unitary transition from
 the $np\to np\,(\pi-\theta)$
 to the $np\to pn\,(\theta)$ elastic representation.

 \section{\boldmath Theoretical approach for $R_{dp}$ and $r^{\rm nfl/fl}$}
 The observable $R_{dp}$ is the ratio of the
 quasi-elastic $nd\to p\,(nn)$ differential cross section to the
 free $np\to pn$ charge exchange one (also named as $CEX$)
\begin{equation}
 R_{dp}=
 \frac{d{\sigma}/dt_{\,\,nd\to
 p\,(nn)}}{d{\sigma}/dt_{CEX}}\,.
\end{equation}
 Following the theory in
 \cite{Dean-1,Dean-2,Luboshitz to Dean},
 where the duration of $nd$ collision is much
 smaller than the characteristic motion period of deuteron
 nucleons, the $nd\to p\,(nn)$ quasi-elastic reaction can be expressed
 within the framework
 of impulse approximation  by the Dean formula
\begin{equation}\label{Dean}
 \frac{d{\sigma}}{dt}_{nd\to p\,(nn)}=
 (1-F(t))\frac{d{\sigma}}{dt}^{\rm Non-Flip}_{CEX}\,
 +(1-\frac{1}{3}F(t))\frac{d{\sigma}}{dt}^{\rm Flip}_{CEX}\,.
\end{equation}
 Here $F(t)$ is the deuteron form-factor which equals one for the
 forward direction, and when the scattering angle $\theta$
 approaches zero,
 the first term on the right-hand of (\ref{Dean}) vanishes
\begin{equation}\label{Dean-simpl}
 \frac{d{\sigma}}{dt}_{nd\to p\,(nn)\,(0)}=
 \frac{2}{3}\frac{d{\sigma}}{dt}^{\rm Flip}_{CEX\,(0)}\,.
\end{equation}
 Note that this simplification is not possible
 if we take the elastic backward reaction $np\to np$
 instead of the charge exchange forward one,
 because if the difference of masses $M_n$ and $M_p$ is neglecting,
 the four-momentum transfer $t$ will be defined as $-4P_{\rm CM}^2$
 and the form-factor $F(t)$ will not equal to one.
 The similar replacement could be justified if both $np$\,-\,elastic
 scattering representations ($np\to np$ backward or
 $np\to pn$ forward) are absolutely identical
 together with their Flip and Non-Flip parts.
 However, this hypothesis is not valid,
 as will be shown in the next section (see also \cite{LL 1,LL 2}).
 Moreover according to the source \cite{Dean-2} the formula (\ref{Dean})
 is defined using the representation of the charge exchange process
 as a ``generalization of the result found originally
 for $K^+d\to K^0pp$\, by Lee \cite{Lee}".
 The author of this work told also that:
 ``For the non-charge-exchange reaction, however, no such simple result follows".

\vskip 3mm
 For $R_{dp}(0)$ and $r^{\rm nfl/fl}_{CEX\,(0)}$ we have
\begin{equation}\label{Rdp&nfl/fl}
 R_{dp}(0)=\frac{2}{3}
 \frac{\frac{d{\sigma}}{dt}^{\rm Flip}_{CEX\,(0)}}{\frac{d{\sigma}}{dt}_{CEX\,(0)}}
 =\frac{2}{3}\frac{1}{1+r^{\rm nfl/fl}_{CEX\,(0)}}
 \,; \qquad
 r^{\rm nfl/fl}_{CEX\,(0)}=\frac{2}{3}\,\frac{1}{R_{dp}(0)}-1\,.
\end{equation}
 Thus, the deuteron as an amplitude filter can be used in the
 measurement of $R_{dp}$ for defining the Flip and Non-Flip
 parts of the $np\to pn$ process, i.e. for observing
 spin effects in the $np$ interaction
 even without the beam and target polarizations.

 \section{\boldmath Transition from the $np\to np\,(\pi-\theta)$
 to the $np\to pn\,(\theta)$ reaction}
 Within the framework of isotopic invariance the nucleon-nucleon scattering
 matrix is
\begin{equation}
 M(k',k)=M_0(k',k)\frac{1-\hat{\tau}_1\hat{\tau}_2}{4} +
 M_1(k',k)\frac{3+\hat{\tau}_1\hat{\tau}_2}{4}\,.
\end{equation}
 Here $\hat{\tau}_1$ and $\hat{\tau}_2$ are the isotopic Pauli
 operators of nucleons, $k$ and $k'$ are the unit vectors of
 the initial and final relative momenta and the matrices $M_0$ and
 $M_1$ describe the $NN$ scattering for the isotopic spin $T=0$ and
 $T=1$ respectively.
 For the $np\to np$ and $np\to pn$ elastic reactions
 at the same angle $\theta$ it can be written
\begin{equation}
 <np|M|np>=\frac{1}{2}(M_1+M_0)
  \qquad \qquad
 <np|M|pn>=\frac{1}{2}(M_1-M_0)
\end{equation}
 With the Pauli spin operators $\hat{\sigma}_1$ and
 $\hat{\sigma}_2$ the scattering matrix $M(k',k)$
 can be expressed in the
 Goldberger--Watson amplitude representation
 \cite{Goldberger,Goldberger-Watson}
\begin{equation}\label{Goldberger-Watson}
 M_T(k',k)=a_T+
 b_T(\hat{\sigma}_1n)(\hat{\sigma}_2n)+
 c_T(\hat{\sigma}_1n+\hat{\sigma}_2n)+
 e_T(\hat{\sigma}_1m)(\hat{\sigma}_2m)+
 f_T(\hat{\sigma}_1l)(\hat{\sigma}_2l)\,.
\end{equation}
 Here $(a,\,b,\,c,\,e,\,f)$ are the complex functions of the
 interacting particle energy and the variable $(k\cdot{k'})=\cos\theta$, the
 index $T$ equals the value of the isotopic spin, and the basic vectors
 are defined as $n=\frac{k\times k'}{|k\times k'|}$, $m=\frac{k-k'}{|k-k'|}$
 and $l=\frac{k+k'}{|k+k'|}$. The Goldberger--Watson formalism is very
 suitable for the separation of elastic scattering into the Flip and
 Non-Flip parts because the amplitude $a_T$ does not have operator
 term and it is Non-Flip by definition
\begin{equation}\label{diff.cross}
\frac{d\sigma}{dt}^{\rm Non-Flip}=|a|^2 \qquad
 {\rm {and}} \qquad
\frac{d\sigma}{dt}=|a|^2+|b|^2+2|c|^2+|e|^2+|f|^2\,.
\end{equation}
 The Wolfenstein formalism\footnote{The vector $m$ in \cite{Wolfenstein} defined
 as $m_w=(k'-k)/|k-k'|$. Therefore the Wolfenstein $(n,m_w,l)$
 basic is left-hand in comparison with the Goldberger-Watson
 definition: $m_w=-m_g$. However the signs of amplitudes will not change
 by the means of bilinear form of operator
 $(\hat{\sigma}_1m)(\hat{\sigma}_2m)$.
 Hereinafter we shall use the right-hand $(n,m,l)$ basis only.}
 \cite{Wolfenstein,First Veritas,Second Veritas}
 allows dividing the matrix
 $M(k',k)$ into the spin-singlet and spin-triplet parts using the
 spin projection operators
 $\hat{S}=\frac{1}{4}(1-\hat{\sigma}_1\hat{\sigma}_2)$ and
 $\hat{T}=\frac{1}{4}(3+\hat{\sigma}_1\hat{\sigma}_2)$
\begin{eqnarray}\label{Wolfenstein}
 M_T(k',k)& = &B_T\hat{S}+
 [\;C_T(\hat{\sigma}_1n+\hat{\sigma}_2n)+
 \frac{1}{2}\,G_T((\hat{\sigma}_1m)(\hat{\sigma}_2m)+
 (\hat{\sigma}_1l)(\hat{\sigma}_2l))+           \nonumber\\
 &   &+\frac{1}{2}\,H_T((\hat{\sigma}_1m)(\hat{\sigma}_2m)-
 (\hat{\sigma}_1l)(\hat{\sigma}_2l))+
 N_T(\hat{\sigma}_1n)(\hat{\sigma}_2n)\;]\,\hat{T}\,.
 \end{eqnarray}
 $B_T$ is the spin-singlet amplitude and the others
 are the spin-triplet amplitudes. Both matrix representations
 (\ref{Goldberger-Watson}) and (\ref{Wolfenstein})
 are related by the linear transitions
\begin{eqnarray}\label{linear-eqs-backward}
 a_T=\frac{1}{4}\,(B_T+G_T+N_T),
 \qquad
 b_T=\frac{1}{4}\,(3N_T-B_T-G_T),
 \qquad
 c_T=C_T\nonumber\\
  e_T=\frac{1}{4}\,(G_T+2H_T-B_T-N_T),
 \qquad
 f_T=\frac{1}{4}\,(G_T-2H_T-B_T-N_T)\,.
\end{eqnarray}
 Let us to quote the works \cite{First Veritas,Second Veritas}:
 ``The requirement of
 antisymmetry of the final wave function $M(k',k)\cdot\chi_S\cdot\chi_T$
 ($\chi_S$ and $\chi_T$ are the spin and isotopic functions of the initial
 state) relative to the total permutation, including the permutation
 of the vector ($k'\to -k'$), permutation of the spin and isotopic
 variables does not change the signs of the amplitudes
 $B_1(\theta)$, $C_1(\theta)$, $H_1(\theta)$, $G_0(\theta)$
 and $N_0(\theta)$ after the turn $\theta\to (\pi-\theta)$, but the
 amplitudes $B_0(\theta)$, $C_0(\theta)$, $H_0(\theta)$,
 $G_1(\theta)$ and $N_1(\theta)$ become inverse".
 It is accepted as the symmetry properties of these amplitudes (Table~1)
 \vspace {-5mm}
\begin{table}[!ht]
 \caption{Symmetry properties of the Wolfenstein amplitudes}
 \begin{center}
 \begin{tabular}{|c|c|}
 \hline
 \phantom{\Large{I}} $T ~=~ 0$
 \phantom{\Large{I}}  &
 \phantom{\Large{I}} $T ~=~ 1$
 \phantom{\Large{I}}   \\
 \hline 
  \phantom{\Large{I}} $B_0(\theta) ~=~ -\,B_0(\pi-\theta)$
  \phantom{\Large{I}}  &
  \phantom{\Large{I}} $B_1(\theta) ~=~ +\,B_1(\pi-\theta)$
  \phantom{\Large{I}}  \\
 \hline
  \phantom{\Large{I}} $C_0(\theta) ~=~ -\,C_0(\pi-\theta)$
  \phantom{\Large{I}}  &  $C_1(\theta) ~=~ +\,C_1(\pi-\theta)$   \\
 \hline
  \phantom{\Large{I}} $H_0(\theta) ~=~ -\,H_0(\pi-\theta)$
  \phantom{\Large{I}}  &  $H_1(\theta) ~=~ +\,H_1(\pi-\theta)$   \\
 \hline
  \phantom{\Large{I}} $G_0(\theta) ~=~ +\,G_0(\pi-\theta)$
  \phantom{\Large{I}}  &  $G_1(\theta) ~=~ -\,G_1(\pi-\theta)$   \\
 \hline
  \phantom{\Large{I}} $N_0(\theta) ~=~ +\,N_0(\pi-\theta)$
  \phantom{\Large{I}}  &  $N_1(\theta) ~=~ -\,N_1(\pi-\theta)$   \\
 \hline
 \end{tabular}
 \end{center}
\end{table}
 \vspace {-5mm}\\
 This rule (Table~1) and
 the symbolical addition $M_1^{CEX}=M_1$, $M_0^{CEX}=-M_0$
 allow the new $np\to pn\,(\theta)$
 charge exchange\footnote{Now each
 of the charge exchange full amplitudes
 $Amp^{CEX}$ is the half-sum of the
 new defined pure isotopic amplitudes
 $Amp^{CEX}_{1}$ and $Amp^{CEX}_{0}$:
 $Amp^{CEX}=1/2\,[Amp^{CEX}_{1}+Amp^{CEX}_{0}]$.}
 forward (Goldberger--Watson) amplitudes to be obtain
 from (\ref{linear-eqs-backward})
 via the old $np\to np\,(\pi-\theta)$ elastic backward
 (Wolfenstein) amplitudes
\begin{eqnarray}\label{linear-eqs-CEX}
 a_T^{CEX}=\frac{1}{4}(B_T-G_T-N_T)
 \qquad
 b_T^{CEX}=\frac{1}{4}(G_T-B_T-3N_T)
 \qquad
 c_T^{CEX}=C_T\nonumber\\
 e_T^{CEX}=\frac{1}{4}(N_T+2H_T-B_T-G_T)
 \qquad
 f_T^{CEX}=\frac{1}{4}(N_T-2H_T-B_T-G_T)\,.
\end{eqnarray}
 As can be seen in (\ref{linear-eqs-backward})
 and (\ref{linear-eqs-CEX}) the Non-Flip
 amplitudes $a_T(\pi-\theta)$ and $a_T^{CEX}(\theta)$ are different
 from each other due to the yield of spin-triplet amplitudes $G_T$
 and $N_T$. For all other Flip terms
 (except the $c^{CEX}_T$ and $c_T$) the yields of $G_T$
 and $N_T$ are also inverted. It is not
 difficult to define the direct 
 amplitude transition from the $np$ elastic backward
 to the charge exchange forward. The amplitudes $c_T^{CEX}$
 and $c_T$ are equal, and for others we have
 \vspace {1mm}
\begin{equation}\label{unitary-transition}
 \left(
 \begin{array}{c}
  a_T^{CEX}(\theta) \\[3mm]
  b_T^{CEX}(\theta) \\[3mm]
  e_T^{CEX}(\theta) \\[3mm]
  f_T^{CEX}(\theta)
 \end{array}
 \right) = A\cdot \left(
 \begin{array}{c}
  a_T(\pi-\theta) \\[3mm]
  b_T(\pi-\theta) \\[3mm]
  e_T(\pi-\theta) \\[3mm]
  f_T(\pi-\theta)
 \end{array}
 \right) \,, \quad {\rm where} \quad A= \left(
 \begin{array}{cccc}
  -\frac{1}{2} & -\frac{1}{2} & -\frac{1}{2} & -\frac{1}{2} \\[3mm]
  -\frac{1}{2} & -\frac{1}{2} & +\frac{1}{2} & +\frac{1}{2} \\[3mm]
  -\frac{1}{2} & +\frac{1}{2} & +\frac{1}{2} & -\frac{1}{2} \\[3mm]
  -\frac{1}{2} & +\frac{1}{2} & -\frac{1}{2} & +\frac{1}{2}
 \end{array}
 \right)\,.
\end{equation}\\

 \noindent
 The inverse transition from the $np\to pn$
 forward reaction to the $np\to np$ backward one will be
 equivalent because the matrix $A$
 is symmetric and unitary:
 $A=A^{-1}\Rightarrow |A|=1$.
 The unitary transition (\ref{unitary-transition})
 and definition (\ref{diff.cross})
 give the equivalence of the differential cross sections
 of both $np$ elastic representations
 even if their Non-Flip or Flip parts are different
\begin{equation}\label{equivalent-diff-cross}
 \frac{d\sigma}{dt}\,np\to np\,(\pi-\theta)=
 \frac{d\sigma}{dt}\,np\to pn\,(\theta)\,.
\end{equation}
 According to the properties of the $NN$ amplitudes,
 when the scattering angle $\theta$ approaches zero, the
 additional simplification arises $b(\pi)=f(\pi)$, $b^{CEX}(0)=e^{CEX}(0)$
 and $c(\pi)=c^{CEX}(0)=0$.
 In this case our formulas will
 coincide with the expressions from \cite{LL 1,LL 2}
\begin{equation}\label{unitary-simple}
 \begin{array}{c}
  a^{CEX}(0)=-\frac{1}{2}(a(\pi)+2b(\pi)+e(\pi))\\[4mm]
  b^{CEX}(0)=-\frac{1}{2}(a(\pi)-e(\pi))\\[4mm]
  f^{CEX}(0)=-\frac{1}{2}(a(\pi)-2b(\pi)+e(\pi))
 \end{array} \,.
\end{equation}

 \noindent
 Here all amplitudes are half-sums of pure isotopic ones.
 We can see again the essential distinction of
 the Non-Flip amplitudes $a^{CEX}(0)$ and $a(\pi)$.
 It is very interesting that the formalism
 of $NN$ elastic scattering was created more than 50
 years ago but this issue was revealed only in 2005 year.
 The elegant method of papers \cite{LL 1,LL 2}
 uses the Hermitian operator of spin permutation
 $\hat{P}=\frac{1}{2}(1+\hat{\sigma}_1\hat{\sigma}_2)$
 and relates both scattering matrices
\begin{equation}
 M^{CEX}(k',k)=-\hat{P}\cdot M^{np\to np}(-k',k)\;\;\;.
\end{equation}
 Dividing the matrices into the spin-singlet $SS$ and spin-triplet $ST$
 parts and using the simplest arithmetics
 $\hat{P}\hat{S}=-\hat{S}$
 and $\hat{P}\hat{T}=+\hat{T}$,
 we can easily define
 \begin{equation}\label{asymmetry}
 M^{np\to np}(-k',k)=SS+ST
  \qquad \qquad
 M^{CEX}(k',k)=SS-ST\,.
 \end{equation}
 The inversion of $ST$ amplitudes\footnote{The symmetry
 properties of the Wolfenstein amplitudes (Table~1) can be
 defined directly from (\ref{asymmetry}):
 the amplitude $B$ ($\in{SS}$) is transformed without change of sign;
 $G$ and $N$ ($\in{ST}$) are inverted; the $C$ and $H$
 (belonging also to the $ST$ part) are inverted twice if we
 take into account that after the turn $k'\to -k'$
 the right-hand basic vectors change too:
 $n\to -n$, $m\to l$ and $l\to m$.}
 is the main cause for the difference
 of these two $np$ elastic representations
 and for the discrepancy between their spin structures.

 \section{Experimental results and comparison with PSA solution}
 According to the research program, the Delta-Sigma collaboration
 has successfully fulfilled the measurements
 of the ratio $R_{dp}(0)$ in four data-taking runs
 in 2002--2007.
 Using the liquid $\rm{D}_2/\rm{H}_2$ targets as well as
 the solid $\rm{CD}_2/\rm{CH}_2/C$ complimentary targets
 we obtained the 8 points at energies T$_n= 0.5$--2.0\,GeV
 (see Tab.~2 and Fig.~1).
 Our preliminary results of $R_{dp}(0)$ measurements
 were published in \cite{Strun,Moroz,Moroz Rdp-error}.
 In addition, Delta-Sigma group have determined in 2007 a
 new data point at T$_{\rm kin}=0.55$\,GeV
 to check the consistency with other
 world experimental data at low energies.
 We presented all our points in \cite{Strunov-conf+epj,My-conf+epj}
 and the full description of the data processing
 and the resulting 7 values
 at energies T$_n= 0.5$, 0.8, 1.0, 1.2, 1.4, 1.8 and 2.0\,GeV
 was given in \cite{JINR} and will be published in \cite{Yad.Phys.}.
 The point at 1.7\,GeV is also measured
 for the first time by the Delta-Sigma collaboration,
 but we have some doubts on its quality. It is related with
 the estimation of number of nuclear in the H$_2$ target,
 and the $R_{dp}(0)$ value at 1.7\,GeV
 have a preliminary status for the present.
 All results of the ratio $R_{dp}(0)$ are very close
 to 0.56 and their errors are $\approx5\,\%$.
 Using (\ref{Rdp&nfl/fl}) we calculated the values of the ratio
 $r^{\rm nfl/fl}_{CEX\,(0)}$ between the Non-Flip and Flip
 parts of the $np\to pn$ charge exchange process
 (see Table~2, Fig.~2).
 Our data are in a good agreement
 with the LAMPF \cite{Boner,Bjork} results (see 3 points below 1\,GeV)
 and coincide exactly with the
 JINR \cite{Glagolev-Rdp} point at 1.0\,GeV.
 Other world values of $R_{dp}(0)$
 were taken from \cite{Lehar-points}.

 For comparison of our and other world data
 on $R_{dp}(0)$ and $r^{\rm nfl/fl}_{CEX\,(0)}$
 with the Phase Shift Analysis (PSA)
 we took from the SAID data base the solutions
 FA91 \cite{FA91}, VZ40 \cite{VZ40} and SP07 \cite{SP07}
 for the $np\to np\,(\pi)$ elastic reaction
 and transformed them to the $np\to pn\,(0)$
 charge exchange representation
 using the unitary transition (\ref{unitary-transition}).
 These energy dependencies were calculated using
 (\ref{Rdp&nfl/fl}) and (\ref{diff.cross}).
 As can be seen, the experimental $R_{dp}(0)$
 and estimated $r^{\rm nfl/fl}_{CEX\,(0)}$ data
 are very similar to the PSA solutions,
 and practically coincide with the FA91 one.
 Without the proper unitary transformation this agreement
 disappears (the PSA curve in Fig.~8
 in \cite{Moroz Rdp-error}).

 \begin{table}[!ht]
 \caption{$R_{dp}(0)$ and $r^{\rm nfl/fl}_{CEX\,(0)}$ results
 and their total errors $\varepsilon_{\rm tot}$}
 \begin{center}
 \begin{tabular}{|c|c|c|c|c|c|c|c|c|}
 \hline
 \phantom{\Large{I}} T$_n$\,\,GeV \phantom{\Large{I}}
  &  0.55   &  0.8    &  1.0    &  1.2    &  1.4    &  1.7    &  1.8    &  2.0    \\
 \hline\hline
 \phantom{\Large{I}} $R_{dp}$ \phantom{\Large{I}}
  &  0.589  &  0.554  &  0.553  &  0.551  &  0.576  &  0.565  &  0.568  &  0.564  \\
 \hline
 \phantom{\Large{I}} $\varepsilon_{\rm tot}$ \phantom{\Large{I}}
 &  0.046  &  0.023  &  0.026  &  0.022  &  0.038  &  0.038  &  0.033  &  0.045  \\
 \hline\hline
 \phantom{\Large{I}} $r^{\rm nfl/fl}$ \phantom{\Large{I}}
 &  0.133  &  0.204  &  0.206  &  0.209  &  0.158  &  0.179  &  0.174  &  0.183  \\
 \hline
 \phantom{\Large{I}} $\varepsilon_{\rm tot}$ \phantom{\Large{I}}
 &  0.088  &  0.051  &  0.057  &  0.048  &  0.077  &  0.080  &  0.068  &  0.094  \\
 \hline
 \end{tabular}
 \end{center}
 \end{table}

 \vspace {-2mm}
 \begin{figure}[!ht]  
 \centering
 \scalebox{.39}{\includegraphics{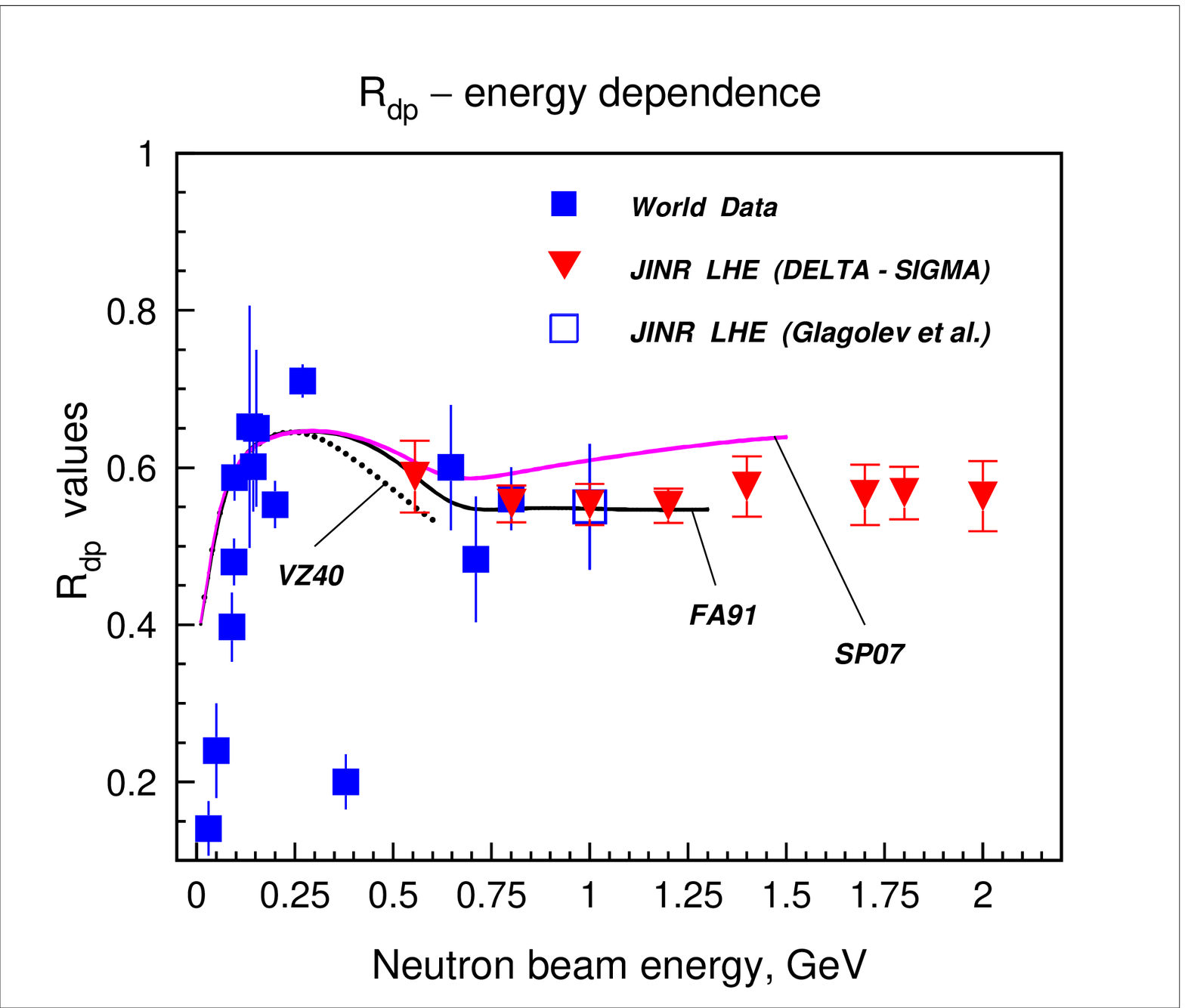}}\hfill
 \scalebox{.39}{\includegraphics{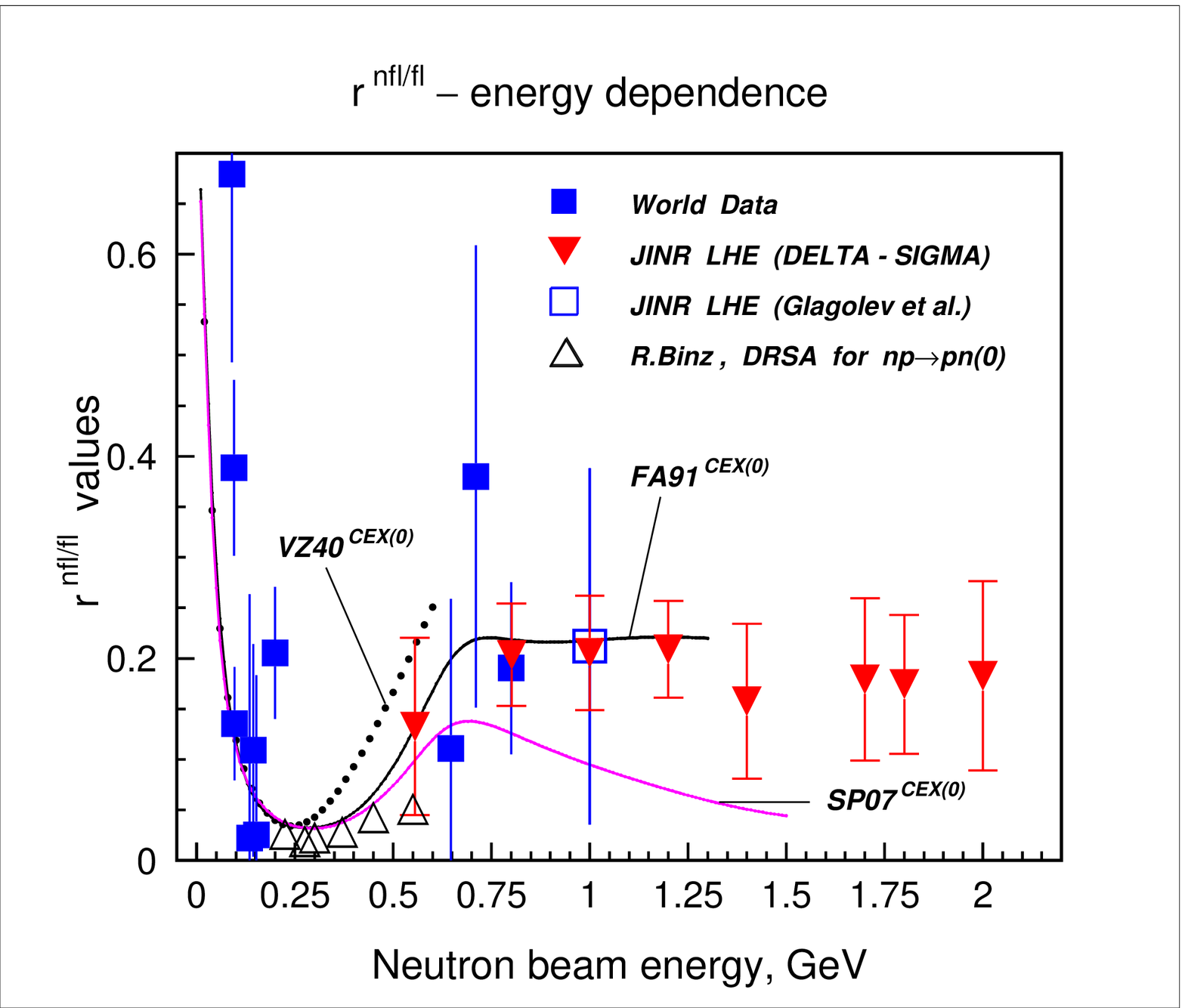}}
 \vspace{2mm}
 \caption{{Energy dependence of the ratio $R_{dp}\,(0)$
 between the yields of the $nd\to p\,(nn)$ quasi elastic
 and $np\to pn$ elastic
 charge exchange reactions. The PSA solutions
 VZ40, FA91 and SP07 were taken from the SAID data base
 as amplitudes for the $np$ backward reaction,
 transformed to the charge exchange by the
 unitary transition (\ref{unitary-transition}), and the
 $R_{dp}\,(0)$ curves are calculated using (\ref{Rdp&nfl/fl}).}}
 \caption{{Energy dependence of the ratio $r^{\rm nfl/fl}_{CEX(0)}$
 between the Non-Flip and Flip parts of the $np\to pn$
 charge exchange elastic process. Our and others
 world points were obtained
 directly from the $R_{dp}\,(0)$ data using (\ref{Rdp&nfl/fl}).
 The PSA solutions are transformed
 by (\ref{unitary-transition}).
 The Binz points from \cite{Binz Ph.D.,Binz Helvetica}
 are the results of the DRSA analysis
 for the $np$ elastic backward reaction
 and they were recalculated again using (\ref{unitary-transition}).}}
 \end{figure}

 \section{Conclusion}
 \begin{itemize}
 \item
 The final \cite{Yad.Phys.} and preliminary (at 1,7\,GeV)
 experimental results of defining
 8 points of the ratio $R_{dp}$
 at the zero angle at energies T$_n=0.5$--2.0\,GeV are presented (see Table~2, Fig.~1).
 The existing world experimental
 data at lower energy agree with our points.
 \item
 With formula (\ref{Rdp&nfl/fl}),
 the values of $r^{\rm nfl/fl}_{CEX(0)}$
 are calculated for the charge exchange process
 $np\to pn\,(0)$ (see Table~2, Fig.~2).
 The Non-Flip part is not zero and equals
 $\approx17$\,\% of the differential cross section.
 \item
 The unitary transition from the $np\to np$ elastic
 backward reaction to the charge exchange $np\to pn$ forward
 process is considered and the PSA curves of $R_{dp}\,(0)$ and
 $r^{\rm nfl/fl}_{CEX(0)}$ calculated by this approach describe
 the experimental points well (see Fig.~1 and Fig.~2).
 \end{itemize}

 \section{Acknowledgement}
 {
 We are grateful to V.\,L.~Luboshitz and Yu.\,N.~Uzikov
 for the theoretical consultations.
 We thank F.\,Lehar for the active discussion though he
 does not agree with the presented method of unitary transformation.
 Our investigations were supported by the
 Russian Foundation for Basic Research,
 projects {\it No}\, 02-02-17129 and {\it No}\, 07-02-01025.}


\end{document}